\documentclass[fleqn,12pt,twoside]{article}
\usepackage[headings]{espcrc1}

\readRCS
$Id: espcrc1.tex,v 1.2 2004/02/24 11:22:11 spepping Exp $
\usepackage{graphicx,floatflt}
\usepackage[figuresright]{rotating}

\newcommand{\AmS}{{\protect\the\textfont2
  A\kern-.1667em\lower.5ex\hbox{M}\kern-.125emS}}

\title{System and Energy Dependence of Strangeness Production with STAR}

\author{Sevil Salur\address[Yale]{
Physics Department, Yale University, P.O. Box 208120, New Haven,
CT, 06520, USA }        (for the STAR\footnote[2]{For the full
list of STAR
        authors and acknowledgments, see appendix `Collaborations' of this volume. } Collaboration)}

\runauthor{S. Salur}

\begin{document}

\maketitle

\begin{abstract}
 The yields and spectra of strange hadrons have each been measured by STAR as a function of centrality
 in $\rm \sqrt{s_{NN}}=$ 200 GeV AuAu collisions. By comparing to pp and
 dAu at $\rm \sqrt{s_{NN}}=200$ GeV and in AuAu at $\rm \sqrt{s_{NN}}=62$ GeV the dependence
 on system size and energy is studied. Strange resonances, such as $\Sigma (1385)$ and
$\Lambda (1520)$, are used to examine the dynamical evolution
between production and freeze-out for these systems. Particle
production is investigated by comparison to thermal models,  which
assume a scaling of the yield with $\rm N_{part}$. Our hyperon
measurements in AuAu indicate that $\rm N_{bin}$ may be a more
appropriate scale for the strangeness correlation volume. Thus
canonical suppression can not be simply parameterized with the
geometrical overlap volume but will depend on the individual quark
content of each particle. This theory is tested by comparing the
data from different collision systems and centralities.
\end{abstract}

\section{Introduction}

\begin{figure}[h!]
\begin{minipage}[h]{75mm}
\vspace{-0.33in}
\includegraphics[width=20pc]{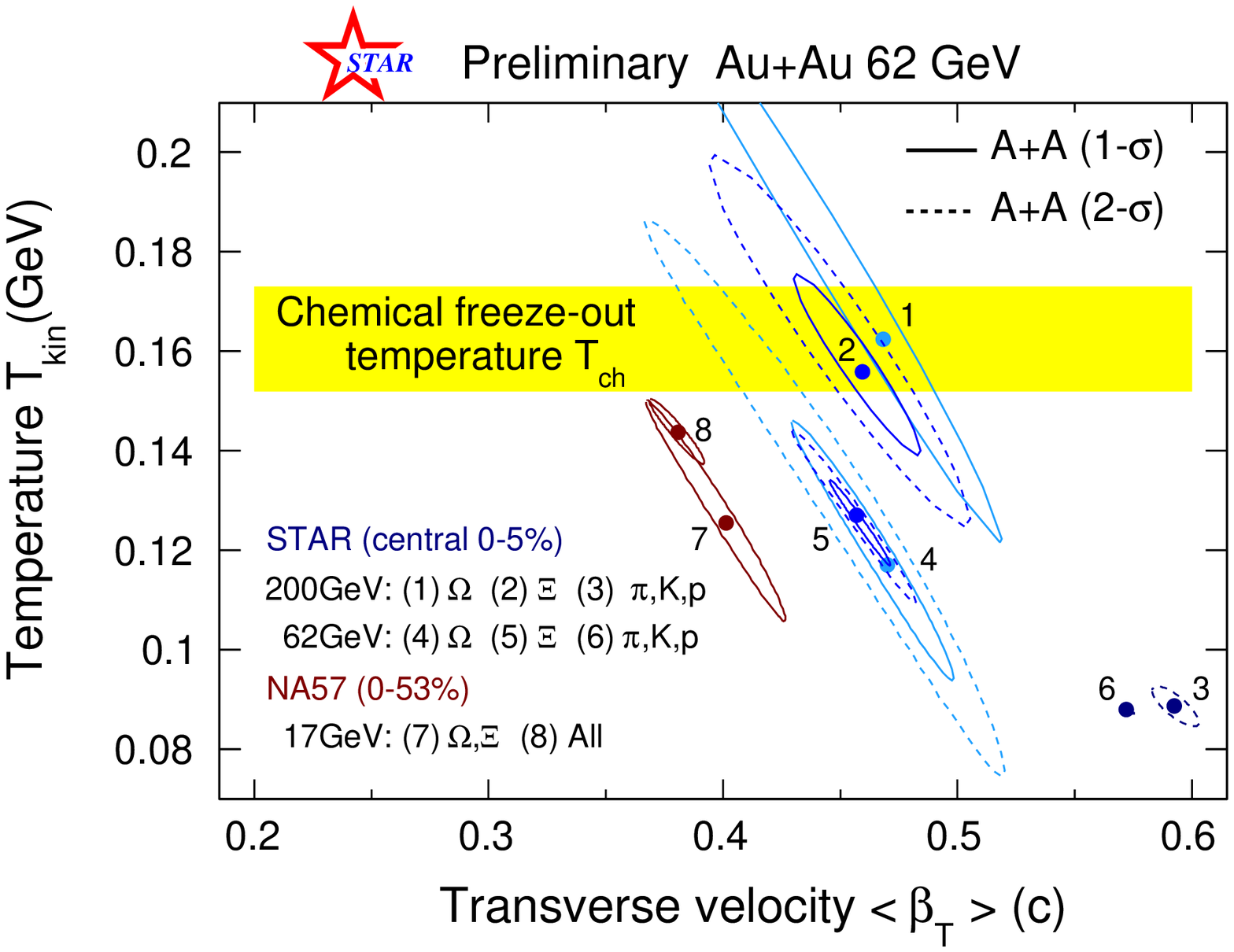}\\
 \label{fig:BW}
\end{minipage}
\hspace{\fill}
\begin{minipage}[h]{75mm}
\vspace{-0.45in}   as at RHIC. The differences between the fit
parameters imply variations in dynamical properties of the
collision energies, while in terms of chemical properties the
results from $\rm \sqrt{s_{NN}}$= 62.4 and 200 GeV collisions are
equivalent \cite{olga}.
 \vspace{-0.3in}\caption{ 1 $\sigma $ and 2 $\sigma $ error contours representing Blast-Wave fits
to particles from the most central AuAu collisions at RHIC  and
PbPb collisions at SPS.  }
\end{minipage}
\vspace{-0.6in}
\end{figure}

RHIC has been run in various configurations of pp, CuCu, dAu, and
AuAu at energies ranging from $\rm \sqrt{s_{NN}}$ =19 to 200 GeV.
These rich data sets, together with STAR's large acceptance,
provide the best opportunity to study strange particles in dense
systems. As the strange quark is the next-lightest after the up
and down quarks and does not exist in the initial colliding
system,  investigation of its production and dynamics may reveal
some of the properties of strongly interacting matter at high
densities.
A hydrodynamically inspired Blast-Wave parametrization with  fit
parameters kinetic temperature $\rm T_{Kin}$ at freeze-out, mean
transverse flow velocity $\langle \beta_{\rm T}\rangle$, and a
normalization factor, is used to fit the data \cite{kolb,jeff}.
The one and two $\sigma$ fit contours of the $\rm T_{Kin}$ and
$\langle \beta_{\rm T}\rangle$ parameters from a Blast-Wave fit to
$\pi$, K, p and strange particles are presented in
Figure~\ref{fig:BW} for $\rm \sqrt{s_{NN}}=$ 17.3, 62.4 and 200
GeV collision energies. The variation of the fit parameters
indicates that the spectral shapes are different for the different
particles. The $\rm T_{Kin}$ parameter is higher (hotter source)
and $\langle\beta_{\rm T}\rangle$ is lower (less flow) for baryons
with higher strange quark content at the same collision energy.
While the $\langle\beta_{\rm T}\rangle$ remains almost unchanged,
the $\rm T_{Kin}$ parameter for the multi-strange baryons is lower
at $\rm \sqrt{s_{NN}}$= 62 GeV than at 200 GeV. This behavior is
different for $\pi$, K, and p where the $\langle\beta_{\rm
T}\rangle$ is larger at 200 GeV though $\rm T_{Kin}$ is the same.
For SPS collisions at $\rm \sqrt{s_{NN}}$=17.3 GeV, the parameters
follow similar trends

\section{System Size Dependence Of Strange Particles}
The energy dependence of $\Lambda$ and $\overline{\Lambda}$ yields
at mid-rapidity from AuAu collisions at RHIC and PbPb collisions
at SPS as a function of $\rm \sqrt{s_{NN}}$ is presented in Figure
2-a. From SPS to RHIC energies, strange baryon production is
approximately constant at mid-rapidity, whereas the
$\overline{\Lambda}$ rises steeply, reaching $80 \%$ of the
$\Lambda$ yield at RHIC top energies. The other hyperons - $\Xi$,
$\Omega$ and $\Sigma(1385)$- follow similar trends. This implies
that at low energies, strange baryon production is dominated by
transport from the colliding system but at RHIC it is dominated by
pair production. Figure~2-b shows the $\overline{\Lambda}/\Lambda$
ratios with respect to the number of participants for the
collisions at SPS and RHIC. Within errors, the
$\overline{\Lambda}/\Lambda$ ratios are nearly independent of the
system size in AuAu collisions of the same energy at RHIC. There
is a decrease in the $\overline{\Lambda}/\Lambda$ ratio moving
from pp to dAu to AuAu.
\begin{figure}[b!]
\vspace{-0.3in}
\begin{minipage}[]{51mm}
\vspace{-0.1in}
\includegraphics[width=12.8pc]{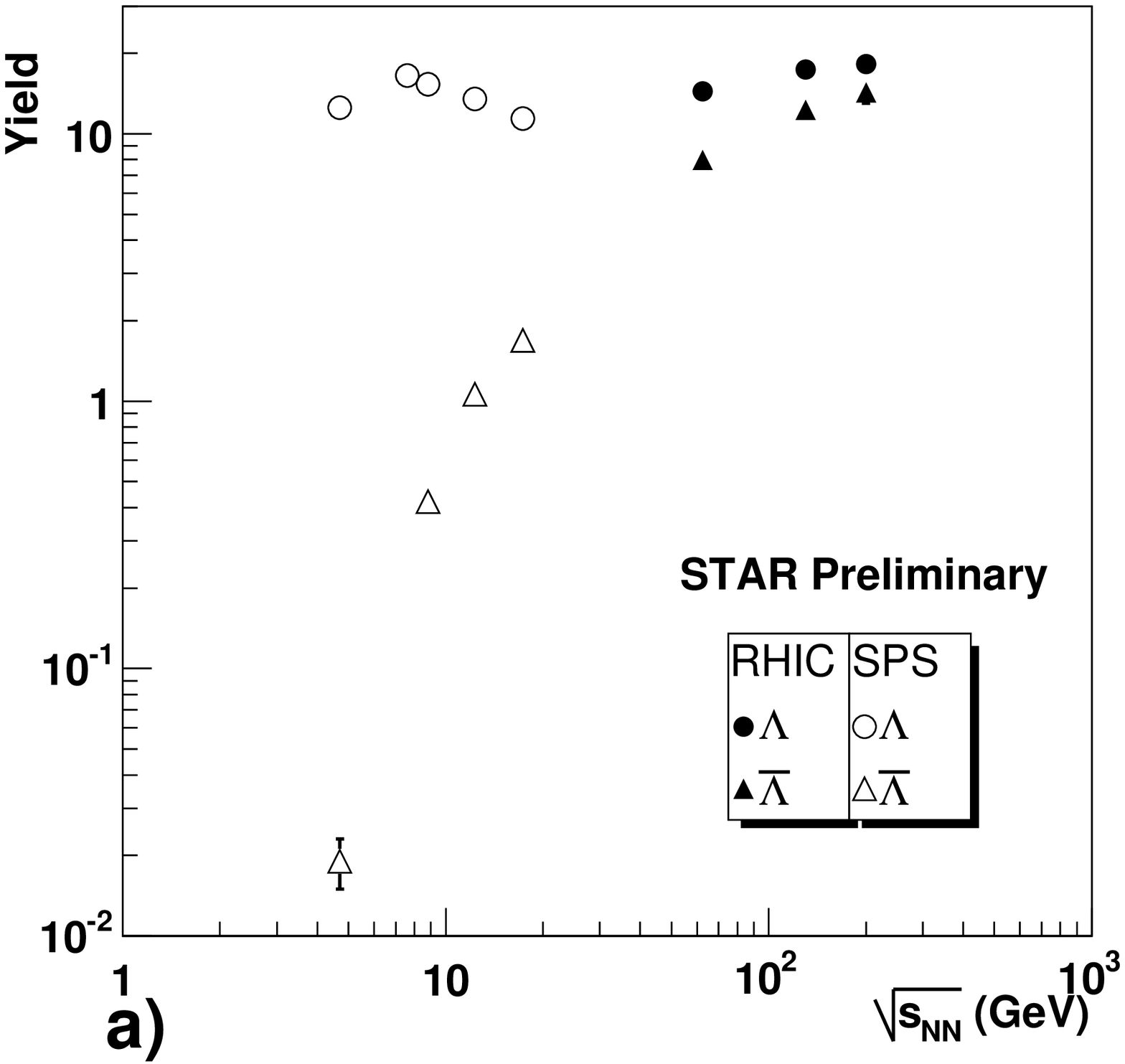}\\
\label{fig:MSRatio}
\end{minipage}
\hspace{-0.01in}
\begin{minipage}[]{52mm}
\includegraphics[width=12.8pc]{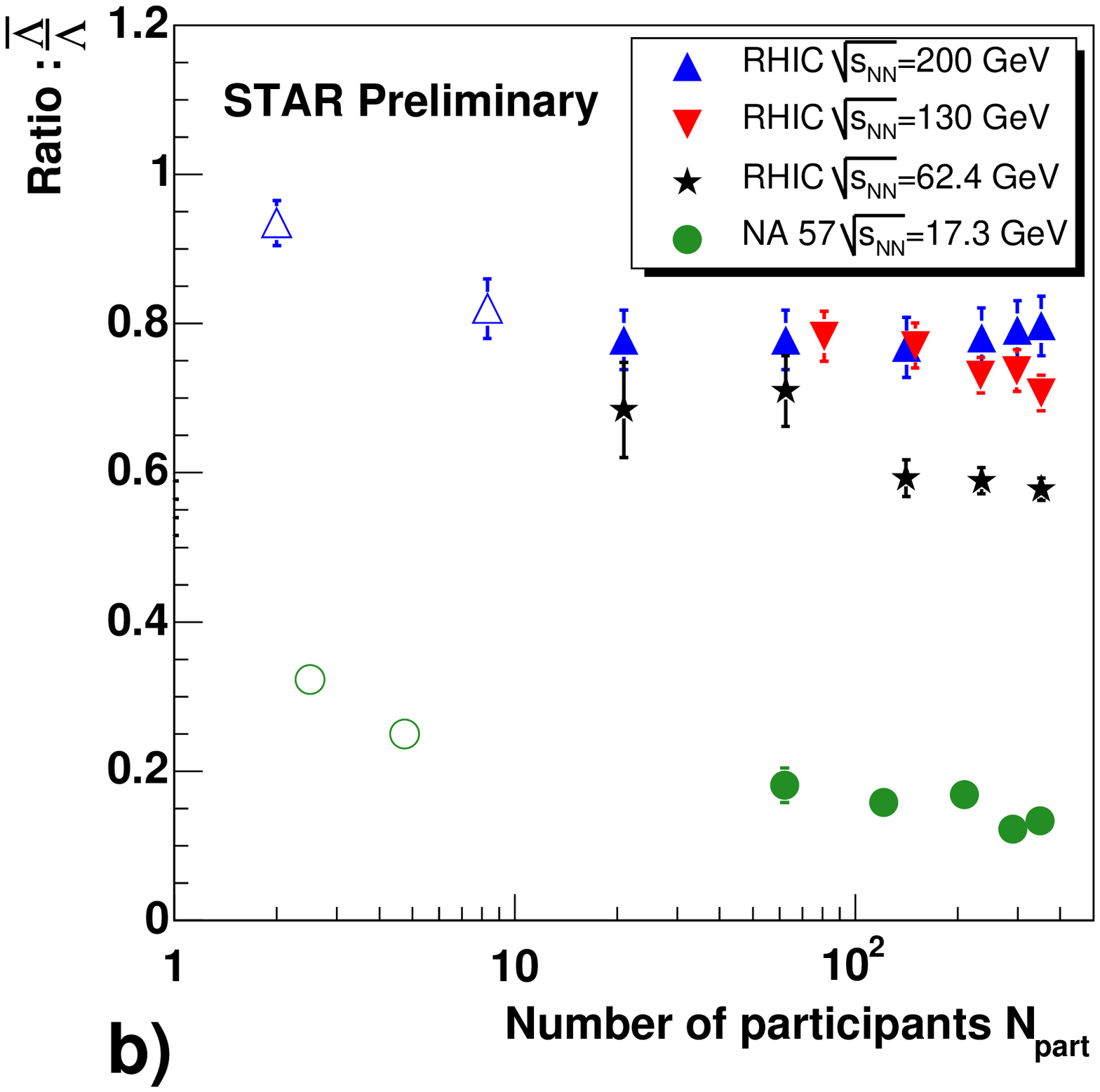}\\
 \label{fig:llratio}
\end{minipage}
\hspace{-0.02in}
\begin{minipage}[]{51mm}
\includegraphics[width=12.8pc]{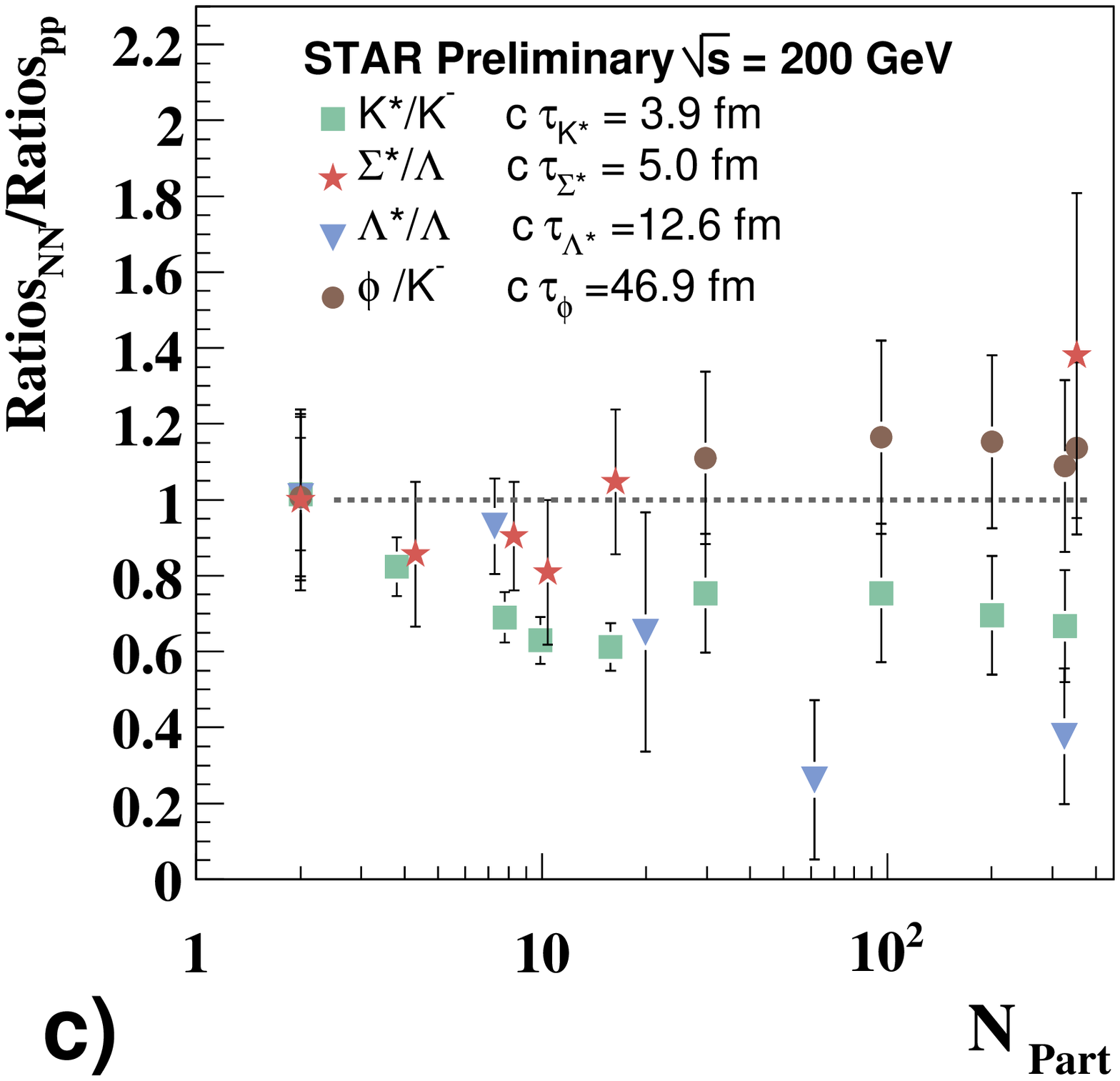}\\
\label{fig:resratio}
\end{minipage}
\vspace{-0.6in} \caption{$\rm (a)$ The $\sqrt{s_{NN}}$ dependence
of $\Lambda$ and $\overline{{\Lambda}}$ yields. $\rm (b)$ The
dependence of $\overline{\Lambda}/\Lambda$ on number of
participants for various collision energies. $\rm (c)$ Resonance
to stable particle ratios  normalized to pp for pp, dAu and AuAu
collisions at $\rm \sqrt{s_{\rm NN}} = $ 200 GeV
\cite{dipak,salur,markert}.} \vspace{-0.1in}
\end{figure}

Figure~2-c shows the ratio of strange resonances to their
corresponding stable particles normalized to their values in pp.
While the $\Sigma(1385)/\Lambda$ ratio is independent of system
size at 200 GeV and is consistent with lower energy pp values,
other ratios such as $\rm K^{*}/K$ and $\Lambda(1520)/\Lambda$
show a slight suppression in AuAu collisions, independent of
centrality. Due to their short lifetimes, the re-scattering of
resonance decay products between chemical and thermal freeze-out
is expected to cause a signal loss. While the observed suppression
of $\rm K^{*}/K$ and $\Lambda(1520)/\Lambda$ corroborates the
re-scattering picture, the lack of suppression of the
$\Sigma(1385)/\Lambda$ ratio implies a recovery mechanism such as
regeneration (e.g. $\Lambda + \pi \rightarrow \Sigma(1385)$). The
 total interaction cross sections with $\pi$
increases from K to p to $\pi$ respectively \cite{PDG}. This
implies that re-scattering of $\rm K^{*}$ decaying into $\pi$ and
K in the medium should be higher than that of $\Lambda(1520)$
decaying into K and p. The shorter lifetime of $\rm K^{*}$
enhances the re-scattering probability. In this scenario, assuming
that the lifetime between chemical and thermal freeze-out is
non-zero, the regeneration cross-section for $\rm K^{*}$ must be
larger than that of $\Lambda(1520)$ due to the smaller suppression
of the $\rm K^{*}$ ratios.

 HBT radii show a linear dependence on $\rm
dN^{1/3}_{ch}/d\eta$, a term related to the final state geometry
through the density at freeze-out \cite{ref:lisa}. If entropy
drives the strangeness yield, results from different collision
energies at the SPS and RHIC should exhibit universal scaling with
entropy. Figure~3 presents the yields of $\bar{\Lambda}$,
$\bar{\Xi}$ in AuAu collisions at RHIC, normalized to yields in pp
and in PbPb collisions at the SPS, normalized to yields in pBe, as
a function of $\rm dN_{ch}/d\eta$ ($\sim$ entropy). Strange yields
in heavy ion collisions, when compared to lighter systems, seem to
universally scale with $\rm dN_{ch}/d\eta$ for SPS and RHIC
energies. It is also predicted that the greater the number of
strange quarks in the particle, the greater the effect of phase
space suppression when modeled with respect to the number of
participants, $\rm N_{part}$ \cite{redlich}. Even though the
expected ordering of the suppression is observed at RHIC, the
~$\bar{\Lambda}$ and ~$\bar{\Xi}$ measurements normalized to their
pp values do not
\begin{figure}[hb!]
\vspace{-0.5in}
\begin{minipage}[h]{45mm}

\includegraphics[width=13.0pc]{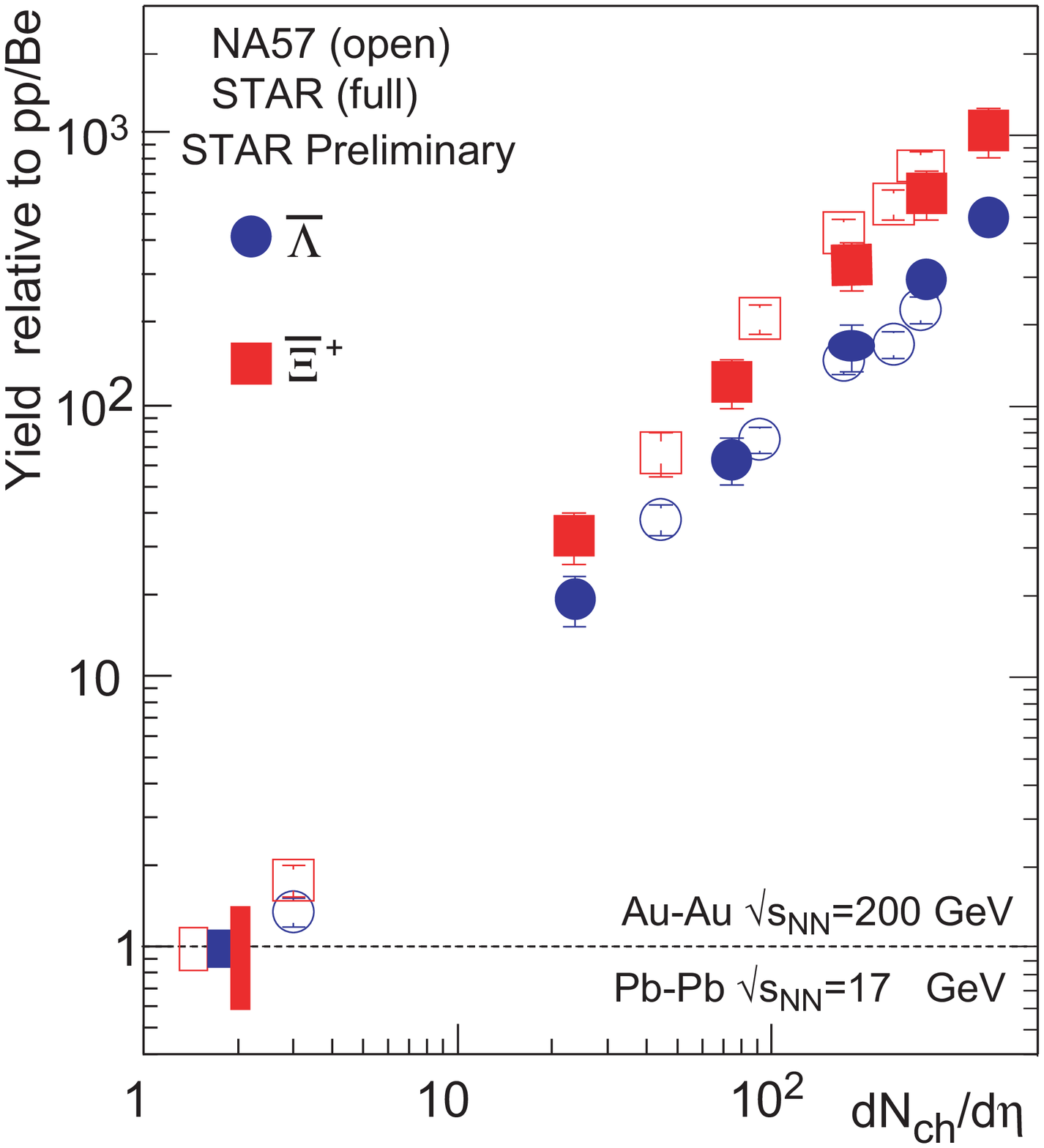}\\

\end{minipage}
\hspace{\fill}
\begin{minipage}[h]{100mm}
 \vspace{-0.15in}flatten  at larger $\rm N_{part}$, in variance to predictions
\cite{helen:sqm}. This might be  because strange particles scale
differently from  non-strange. Particles with only u and d quarks
are already observed to scale with $\rm N_{part}$ while strange
quarks appear to scale better with $\rm N_{bin}$. $\rm N_{part}$
and $\rm N_{bin}$ can be combined to form a correlation volume for
strange particles which depends on the quark content. This
combined scaling seems to represent the strange particles better
than  $\rm N_{part}$ alone.
 \vspace{-0.35in}
 \caption{The dependence on $\rm dN_{ch}/d \eta $ of the $\bar{\Lambda}$ and $\bar{\Xi}$
 yields in AuAu relative to pp at RHIC and in PbPb relative to pBe at SPS energies.
 }
\end{minipage}
 \label{fig:dndeta}
 \vspace{-0.7in}
\end{figure}


 Nuclear modification factors for strange particles in AuAu
 collisions are presented in Figure~4. At higher $\rm p_{T}$,
the ratios exhibit a suppression from binary scaling, attributed
to fast moving partons losing energy as they traverse a dense
medium.
  The $\rm R_{CP}$ from $\rm \sqrt{s_{NN}}=$ 62.4 GeV shows less
  suppression than that at $\rm \sqrt{s_{NN}}=$ 200 GeV; however, the clear differences between baryons
  and mesons still exists \cite{rcp}. This is believed to be due to hadron production through quark coalescence
  at intermediate $\rm p_{T}$  \cite{matt}. For baryons and mesons, the suppression
   sets in at a different $\rm p_{T}$. Motivated by the coalescence picture, Figure~4-b
 shows the $\rm R_{CP}$ ratio vs $\rm p_{T}/n$ for $ \rm \sqrt {s_{NN}}=62.4 $ GeV, where n is the number of
 valence quarks. Thus $\rm p_{T}/n$ is the $\rm p_{T}$ of a quark. The baryon and meson sets
 in at the same quark $\rm p_{T}$, in agreement with the coalescence
 picture. This is also observed for  $\rm \sqrt{s_{NN}}=200 $
 GeV collisions.
\begin{figure}[h!]
\vspace{-0.3in}
\begin{minipage}[t]{51mm}
\includegraphics[width=12.7pc]{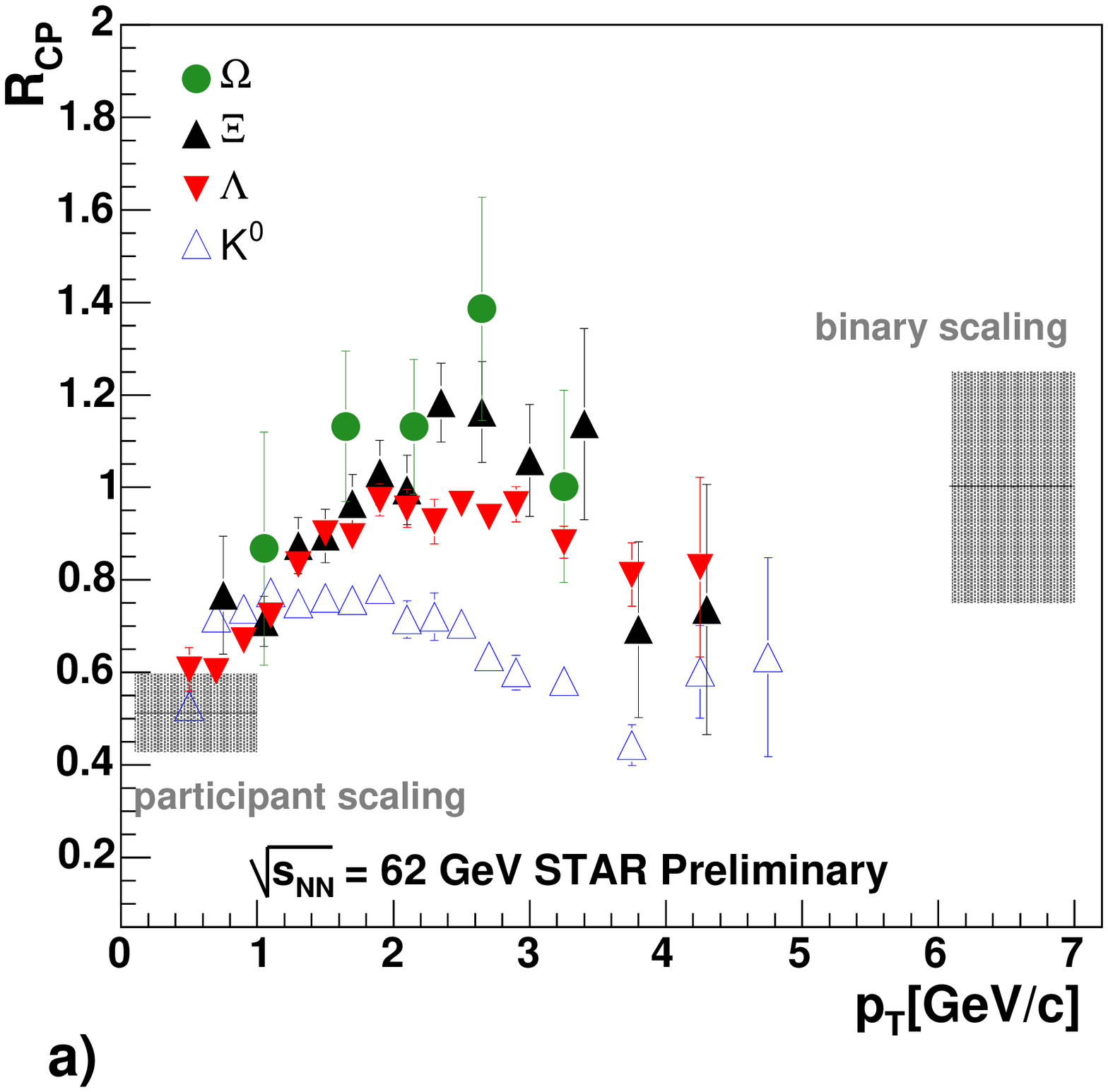}\\
\label{fig:Rcp62}
\end{minipage}
\hspace{-0.01in}
\begin{minipage}[t]{51mm}
\includegraphics[width=12.3pc]{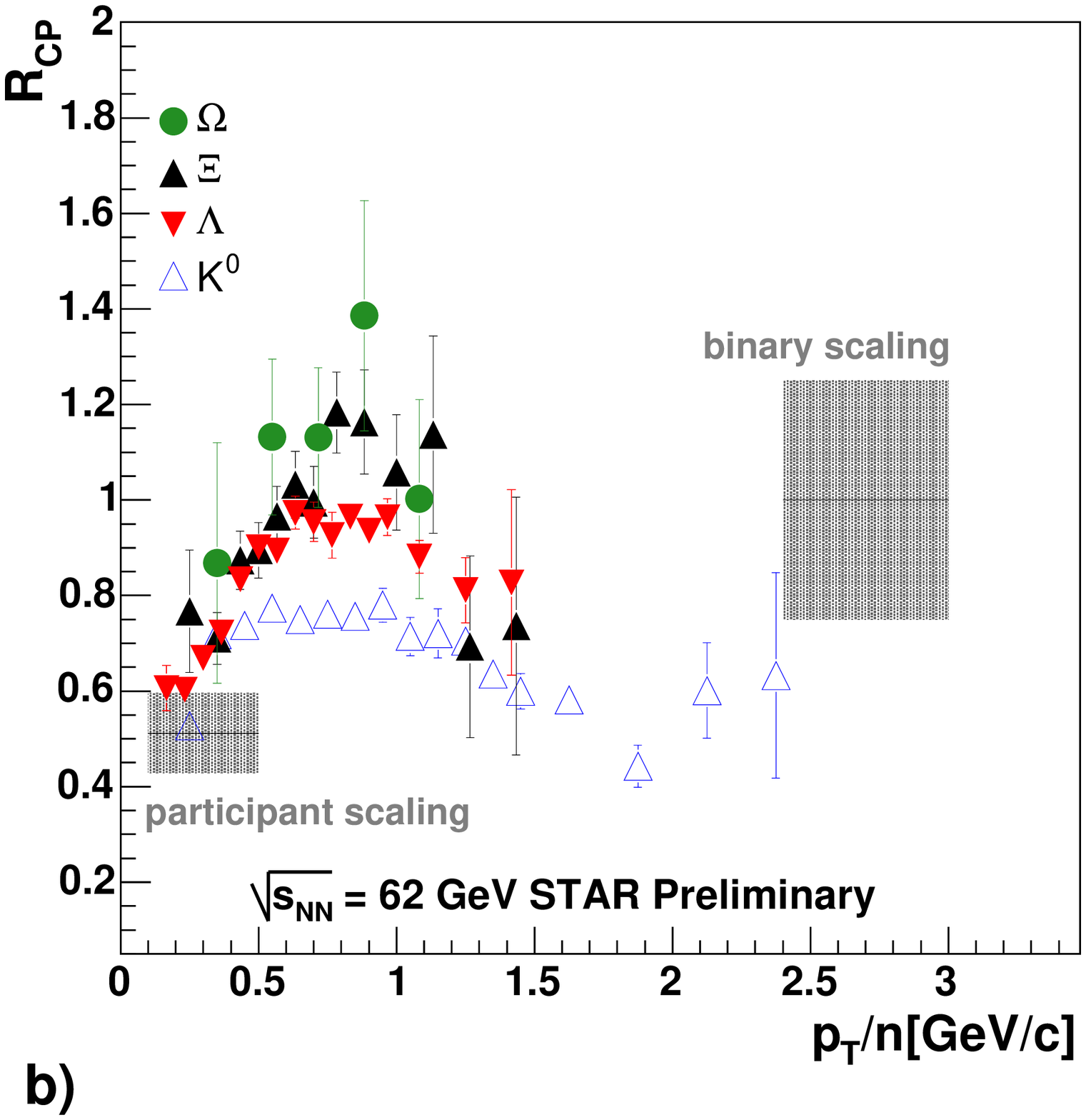}\\
\label{fig:Rcp}
\end{minipage}
\hspace{-0.03in}
\begin{minipage}[t]{51mm}
\includegraphics[width=12.7pc]{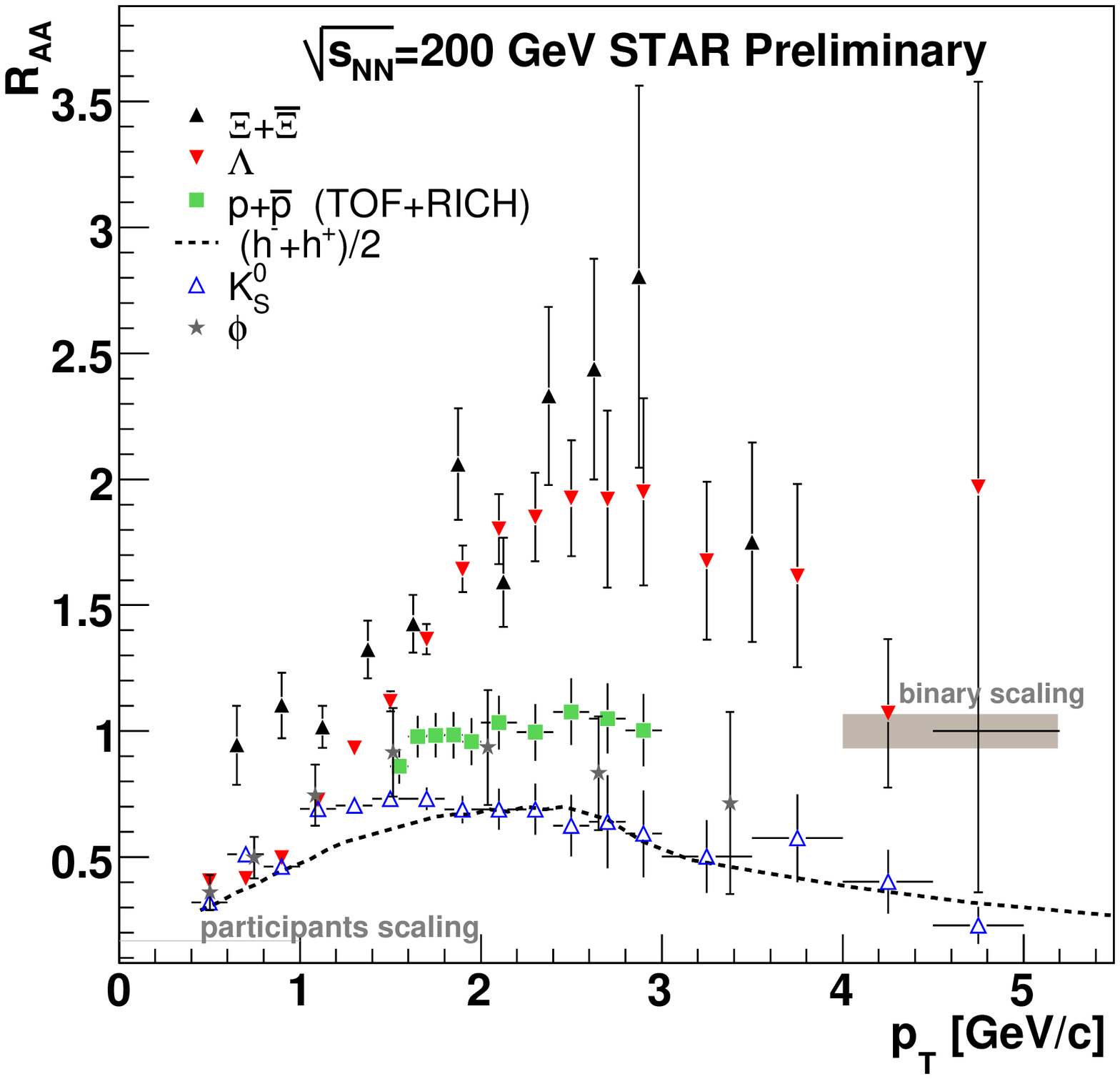}\\
\label{fig:RcpScaled}
\end{minipage}\vspace{-0.55in}
\caption{(a) $\rm R_{CP}$ vs $\rm p_{T}$ at $\rm
\sqrt{s_{NN}}=62.4$ GeV. (b)  $\rm R_{CP}$ vs $\rm p_{T}/n$ at
$\rm \sqrt{s_{NN}}=62.4$ GeV. (c)~$\rm R_{AA}$ at $\rm
\sqrt{s_{NN}}=200$ GeV with respect to $\rm p_{T}$. $\rm R_{CP}$
is calculated from 0-5\% and 40-60\% central AuAu collisions and
$\rm R_{AA}$ is from 0-5\% central AuAu and min-bias pp
collisions.}\vspace{-0.3in}
\end{figure}
The measurement of $\rm R_{AA}$ with respect to $\rm p_{T}$ is
shown in Figure~4-c. While the $\rm R_{AA}$ for mesons ($\rm h^{+}
+ h^{-}$, $\rm K^{0}_{S}$, $\phi$)  is similar to their $\rm
R_{CP}$ values, $\rm R_{AA}$ of strange baryons shows significant
differences (i.e. no suppression). Instead, there is an
enhancement and ordering with strangeness content: the higher the
strangeness content, the higher the $\rm R_{AA}$ measurement in
the intermediate $\rm p_{T}$ region. The difference between yields
in pp and peripheral AuAu may be explained by phase space
(canonical) suppression in the pp data-set although this is
usually attributed
to low $\rm p_{T}$ particles \cite{redlich2}. 

\section{Conclusions}
Dynamical properties of  strange particles can be investigated by
Blast-Wave fits.  The $\langle \beta_{T} \rangle $ increases with
collision energy, implying a higher flow and the freeze-out
temperatures of multi-strange baryons are larger than those of
light mesons, suggesting an earlier freeze-out (e.g. $\rm T (\Xi)
> T(\pi)$). Strange anti-baryon and baryon production are approximately equal
at top RHIC energies.  There is a slight decrease in the
$\overline{\Lambda}/\Lambda$ ratio from pp to dAu to AuAu,
indicating that baryon number transport is almost independent of
system size. Resonance yields require both re-scattering and
regeneration mechanisms for $\Delta t > 0$ between chemical and
thermal freeze-out to describe the suppression in some of their
yields. Strange quarks appear to scale better with the number of
hard processes ($\rm N_{bin}$), while light quarks scale with $\rm
N_{part}$ and strange particle yields seem to universally scale
with $\rm dN_{ch}/d\eta$ for SPS and RHIC. The meson/baryon
separation of the nuclear modification factors also exists in $\rm
\sqrt{s_{NN}}=62$ GeV collisions which can be explained in a
coalescence picture. The $\rm R_{AA}$ of strange baryons behave
differently from their $\rm R_{CP}$. Canonical suppression in pp
might explain the observed difference, though it is a surprise
that this effect extends to intermediate $\rm p_{T}$.

{

}
\end{document}